\documentclass[twocolumn,showpacs,amsmath,amssymb,prl, superscriptaddress]{revtex4} 
\usepackage{graphicx}
\usepackage{dcolumn}
\usepackage{bm}
\usepackage{color}

\begin{document}


\title{Magnetoresistance of disordered graphene: from low to high temperatures}

\author{B. Jabakhanji}
\affiliation{Universit\'e Montpellier 2-CNRS, Laboratoire Charles Coulomb UMR 5221, F-34095, Montpellier, France}
\affiliation{American University of the Middle East (AUM), College of Engineering and Technology, Egaila, Kuwait}
\author{D. Kazazis}
\affiliation{CNRS-Laboratoire de Photonique et de Nanostructures, Route de Nozay, 91460 Marcoussis, France}
\author{W. Desrat}
\affiliation{Universit\'e Montpellier 2-CNRS, Laboratoire Charles Coulomb UMR 5221, F-34095, Montpellier, France}
\author{A. Michon}
\author{M. Portail}
\affiliation{CRHEA-CNRS UPR 10, rue B. Gr\'egory, Parc de Sophia Antipolis, 06560 Valbonne, France}
\author{B. Jouault}
\affiliation{Universit\'e Montpellier 2-CNRS, Laboratoire Charles Coulomb UMR 5221, F-34095, Montpellier, France}

\begin{abstract}
We present the magnetoresistance (MR) of highly doped monolayer graphene layers grown by chemical vapor deposition on 6H-SiC. 
The magnetotransport studies are performed  on a large temperature range, from $T$ = 1.7 K up to room temperature.
The MR exhibits a maximum in the temperature range $120-240$ K. The maximum is observed at intermediate magnetic fields ($B=2-6$ T), in between the weak localization and the Shubnikov-de Haas regimes.
It results from the competition of two mechanisms. First, the low field magnetoresistance increases continuously with $T$ and has a purely classical origin. This positive MR is induced by thermal averaging and finds its physical origin in the energy dependence of the mobility around the Fermi energy. Second, the high field negative MR originates from the electron-electron interaction (EEI). The transition from the diffusive to the ballistic regime is observed. The amplitude of the EEI correction points towards the coexistence of both long and short range disorder in these samples. 

\end{abstract}

\maketitle

\section{Introduction}

Graphene is a newly discovered electronic material which attracts a lot of attention, for both potential applications and unique physical properties~\cite{nobellecture}.
For electronics, the improvement of the graphene quality requires  the identification of  the main sources of scattering which limit the mobility and this is usually done by transport experiments~\cite{Tikhonenko2008,Ponomarenko2009, Monteverde2010, Guignard2012}. 
At intermediate magnetic fields, between the realms of weak localization and quantum Hall effect, recent measurements highlighted the role of the electron-electron interaction (EEI) on the magnetoconductivity.\cite{Kozikov2010, Jouault2011, Jobst2012, Iagallo2013} 
A complete theory of EEI in graphene is still missing, but it is possible to 
use the knowledge accumulated in more conventional two-dimensional systems like thin metal films or semiconductor heterostructures, for which EEI has been theoretically and experimentally studied over more than three decades~\cite{Altshuler1985, Bergmann1984,Mirlin2001, Polyakov2001, Minkov2003, Goh2008}. 
The quantum correction due to the EEI  differs at low and high temperatures.
When the effective interaction time, $\hbar/k_B T$, is larger than 
the transport time $\tau$,
the electrons experience many collisions during their interaction: they are in a diffusive regime and 
the correction of the zero-field conductivity follows a logarithmic dependence on the temperature. When $k_BT\tau/\hbar\gg1$, 
electrons collide at most with one impurity.
This ballistic regime leads to linear-in-$T$ corrections to the zero-field conductivity. 
 
A first complete theory, unifying both regimes, was developed~\cite{Zala2001}
but is only valid if impurities can be considered as point-like scatterers.
This condition is satisfied in Si metal-oxide-semiconductor field effect transistor (MOSFET)~\cite{Pudalov2003, Shashkin2002} and in some semiconductor heterostructures, where carrier scattering is dominated by background impurities~\cite{Coleridge2002, Proskuryakov2002}. 
In graphene, the validity of this theory is questionable as the dominant scattering depends on both the graphene quality and the characteristics of the environment.~\cite{peres}
In this paper, we will rely on a more recent theory~\cite{Gornyi2004} which predicts the EEI correction at all temperatures for both short and long ranges disorder.
Besides, EEI in graphene is specifically sensitive to the type of disorder, both in the diffusive~\cite{Kozikov2010, Jobst2012} and ballistic regimes with unusual temperature dependencies. In particular, in the ballistic regime, EEI depends on the impurity type~\cite{Cheianov2006} and can give indications on the microscopic nature of disorder in graphene.
Up to now, most of the EEI measurements in graphene have focused on the diffusive regime~\cite{Kozikov2010, Jouault2011, Jobst2012, Iagallo2013}. The systematic study of EEI correction in this material, from the diffusive to the ballistic regime, associated with quantitative and qualitative comparisons with models of disorder, is still lacking. This is the scope of this paper.

\section{Methods}
The samples are large and homogeneous single graphene layers obtained by chemical vapor deposition using propane-hydrogen mixtures.\cite{Michon2010,Michon2013} 
Depending on the growth conditions, 
either the samples are hole-doped and the graphene lies on a hydrogen-passivated SiC surface,
or they are $n$-doped and the graphene lies on a carbon-rich buffer layer.~\cite{riedl09, Jabakhanji2014}. Because of their large size, their good spatial homogeneity, these samples are especially well suited for the analysis of weak localization and EEI.  

After the growth, Raman Spectra are recorded using an Acton spectrometer fitted with a Pylon CCD detector and a 600 grooves/mm grating. The samples
are excited with a 532 nm (2.33 eV) continuous wave frequency doubled Nd:Yag laser through a $\times$100 objective (numerical aperture 0.9). The full width at half maximum of the focused laser spot is about 400 nm.

Electrical measurements are performed on $100$ or 20  $\mu$m-wide Hall bars processed by electron beam lithography. 
The Hall bar geometry has been chosen to minimize the current deflection, and follows the technical guidelines which are in use for metrological measurement of the quantum Hall effect. The distance between adjacent lateral probes corresponds to the sample width. The ohmic contacts are fabricated by metal deposition of Pd/Au with an ultra-thin Ti sublayer. 
The samples are covered by PMMA for additional protection. The magnetoresistivities are measured by lock-in amplifiers with low-frequency currents ($f \sim 10$ Hz, $I=0.01-1$ $\mu$A) in a variable temperature insert from $T=1.7$ K up to room temperature. Magnetic fields in the range 0-8 T are used. 

In total, we investigated five samples, whose parameters are given in Table~\ref{table1}.  Samples S1, S2 and S5 have been done with similar growth procedures. The SiC substrate is their case is hydrogen passivated, and the graphene is $p$-doped.  Samples S3 and S4 have been done with another growth procedure, at a higher temperature. The graphene is $n$-doped and resides on a carbon-rich interface. For details, see Ref.~\cite{Jabakhanji2014}.
This paper presents mainly the results obtained for samples S1 and S2. The data obtained for samples S3-S5 lead to similar conclusions and are discussed in the last section.

\begin{table}[b!]
\begin{center}
  \begin{tabular}{ | l | p{1.1cm} | p{1.1cm} | p{1.1cm} | p{1.1cm} |  p{1.1cm} | }
    \hline
    Sample & $p$ & $\mu$ & $\epsilon_F$   & $\tau$ & $\tau_q$      \\ \hline
    S1 &  6.9  & 2770 & 3500 &   79  &  17              \\ \hline
		S2 &  7.4 & 2560  &3700  &   71  &  14                 \\ \hline
		S3 &  -2.0 & 3000  &1800  &   50 &  18                 \\ \hline
		S4 &  -3.1 & 2300  &2400  &   50 &  14                 \\ \hline
		S5 &  6.2 & 2400  &3400  &   70  &  18                 \\ \hline
  \end{tabular}
\end{center}
\caption{Main parameters of the samples investigated in this work: carrier concentration $p$ (in $10^{12}$cm$^{-2}$), mobility $\mu$(in cm$^2$/Vs), Fermi energy $\epsilon_F$ (in K), transport and quantum scattering times $\tau$ and $\tau_q$ (in fs). All parameters are given for $T$=1.7 K. Sample S4 has a width of 20 $\mu$m. The other samples have a width of 100 $\mu$m.
}
\label{table1}
\end{table}

\section{Results}
\begin{figure}[]
\includegraphics[width=\columnwidth]{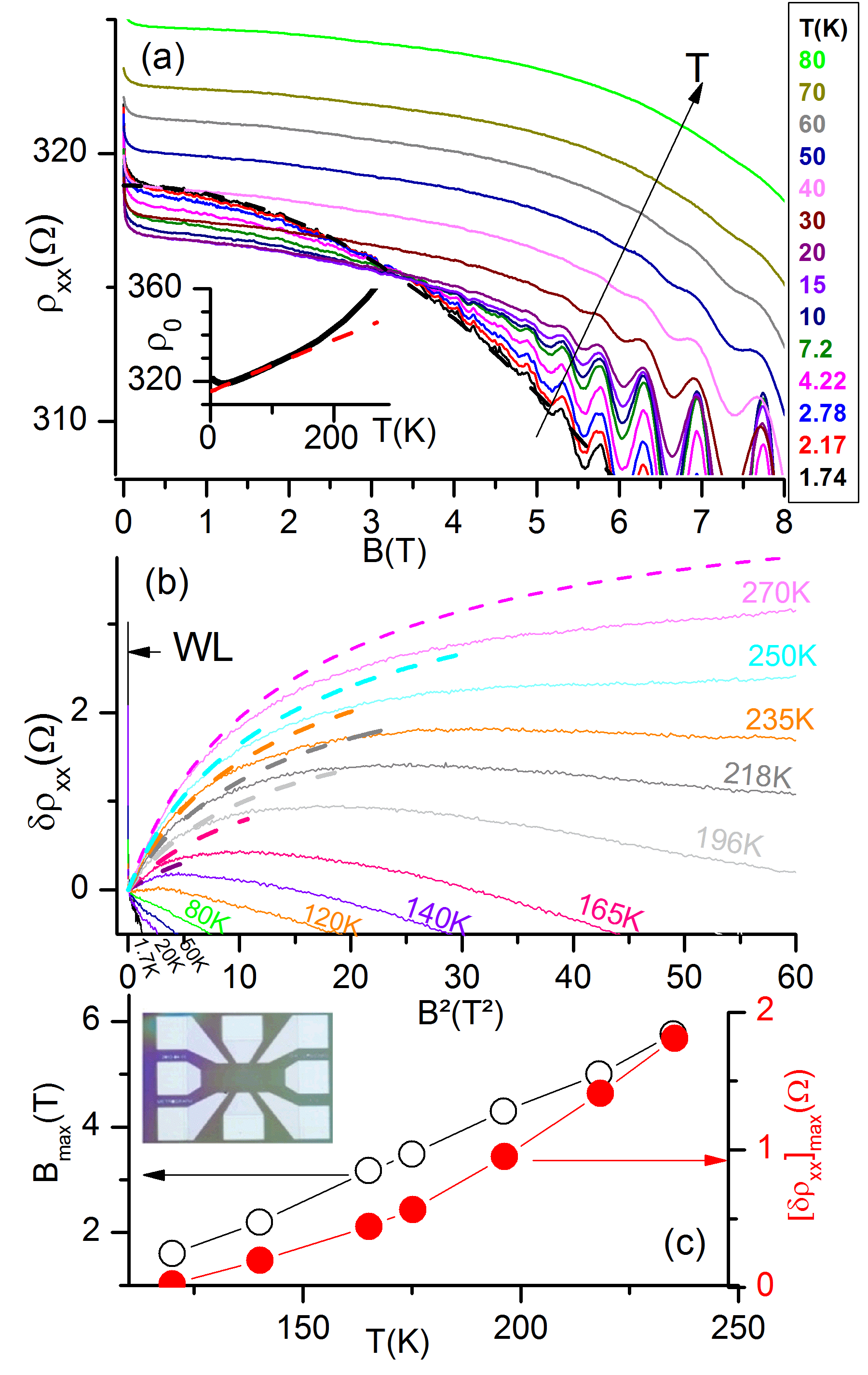}
\caption{ (a) Longitudinal resistivity of sample S1 {\it vs} magnetic field for several temperatures from $T$=1.7 K to 80 K. The dashed line is a guide for the eye which corresponds to a generic parabola and can be compared to the data at $T$= 1.7 K. A weak localization peak is visible below 300 mT. The inset shows the resistivity $\rho_0$ at $B=0$ as a function of the temperature for sample S1 (solid line). The model $\rho_T(B=0)$ is also shown  (red dashed line). (b) $\delta\rho_{xx}$ as a function of $B^2$ from $1.7$ K up to $270$ K. Dashed lines are the expected corrections to the magnetoresistivity $\rho_T(B)-\rho_T(B=0)$ induced by thermal averaging, indicated at 7 temperatures from 140 to 270 K.
(c) Open black and solid red circles are respectively the positions and the amplitudes of the $\rho_{xx}$ maximum vs $T$. Inset: optical view of one of the samples.
}
\label{fig:data}
\end{figure}

The inset of Fig.~\ref{fig:data}(a) shows the zero-field resistivity $\rho_0$ of sample S1. From 300 K down to 30 K, it decreases with $T$. This is related to the slight mobility increase from $2540$ up to $2770$ cm$^{2}$V$^{-1}$s$^{-1}$ between $271$ K and $30$ K, while the hole density remains constant for this temperature range and is equal to $p=6.9\times10^{12}$ cm$^{-2}$. Below $T=30$~K, the resistivity saturates at $\rho \approx 320~\Omega$, which gives a constant carrier mobility $\mu$ = $2770$ cm$^2$V$^{-1}$s$^{-1}$ and a transport scattering time $\tau=79$ fs. 

Figure~\ref{fig:data}(a) shows the longitudinal magnetoresistivity $\rho_{xx}(B)$ for sample S1, measured at several temperatures up to $T$= 80 K.
Above $B=4$ T, Shubnikov-de Haas (SdH) oscillations emerge. The analysis of the damping of the oscillations, as detailed in the Appendix, gives a quantum scattering time of $\tau_q$= 17 fs.
The large ratio $\tau/\tau_q \sim$ 4.5 at $T=1.7$ K reveals the presence of long-range disorder. Either long range scattering governs both  $\tau$ and $\tau_q$, or there is a mixed disorder situation where long and short range scatterings dominate  $\tau_q$ and $\tau$ respectively. In both cases, the presence of long range disorder indicates that Ref.~\cite{Gornyi2004} is appropriate for the analysis of the EEI interaction. 


Below $T=$ 30 K, $\rho_{xx}(B)$ shows the typical features of EEI in the diffusive regime: it is almost perfectly parabolic and its curvature decreases logarithmically with $T$. Moreover, all $\rho_{xx}(B)$ curves taken at different $T$ cross at $\mu B  \sim 1$, as expected by the theory.~\cite{Altshuler1985}
Between 30 K and 80 K, $\rho_{xx}$ remains parabolic but increases with $T$ as phonon scattering is not negligible. Above 80 K, $\rho_{xx}(B)$ is not parabolic anymore and a flat region starts to develop at low fields $\mu B < 1$.~\cite{Gornyi2004,Li2003}

The MR data at higher $T$ have been drawn in Fig. \ref{fig:data}(b) as $\delta\rho_{xx}$ {\it vs} $B^2$, where
$\delta\rho_{xx}= \rho_{xx}(B)- \rho_{xx} (B_{ref})$. The magnetic field reference $B_{ref}$= 300 mT has been chosen to minimize the effects of weak localization which is visible at lower $B$.
 Above 80 K, an unexpected feature appears: the magnetoresistance is positive at low magnetic field, reaches a maximum and then decreases at higher field. The magnetic field position and the amplitude of this resistance maximum are plotted in Fig.~\ref{fig:data}(c). The peak position $B_{max}$ shifts linearly from $B=1.7$ T to $5.6$ T as a function of the temperature with a slope of $0.034$ T/K. This value is very close  to the condition $\hbar\omega_c=k_BT$ which
gives a slope equal to $0.035$ T/K. As $\hbar \omega_c \gg k_B T$ is one of the conditions required (with $\mu B \gg 1$) to observe the EEI correction in the ballistic regime in Ref.~\cite{Gornyi2004}, it is legitimate to attribute the negative MR (NMR) on the high-$B$ side to EEI, but the positive MR (PMR) at low fields also needs an explanation.

Finally, on almost the whole temperature range, a weak localization (WL) peak is also visible at $B <$ 300~mT. 
The attribution of this peak to WL is straightforward, as it gives a correction $\delta\rho_{wl}$ to the resistivity with the expected amplitude and temperature dependence: $\delta \rho_{wl}/\rho_{xx}^2 \sim -(e^2 /\pi h) \ln(T)$.~\cite{Kechedzhi}
At the contrary, there is no sign of weak antilocalization (WAL). While WAL can also give rise to a positive magnetoresistance, both WAL and WL should disappear when $T$ increases. This was experimentally verified by Tikhonenko et al.~\cite{Tikhonenko2008} who observed that at $T > 200$K WAL disappears due to rapid dephasing of the electron trajectories. In our work, the amplitude of the positive magnetoresistance increases continuously with the temperature and consequently cannot be attributed to WAL.

\begin{figure}[]
\includegraphics[width=\columnwidth]{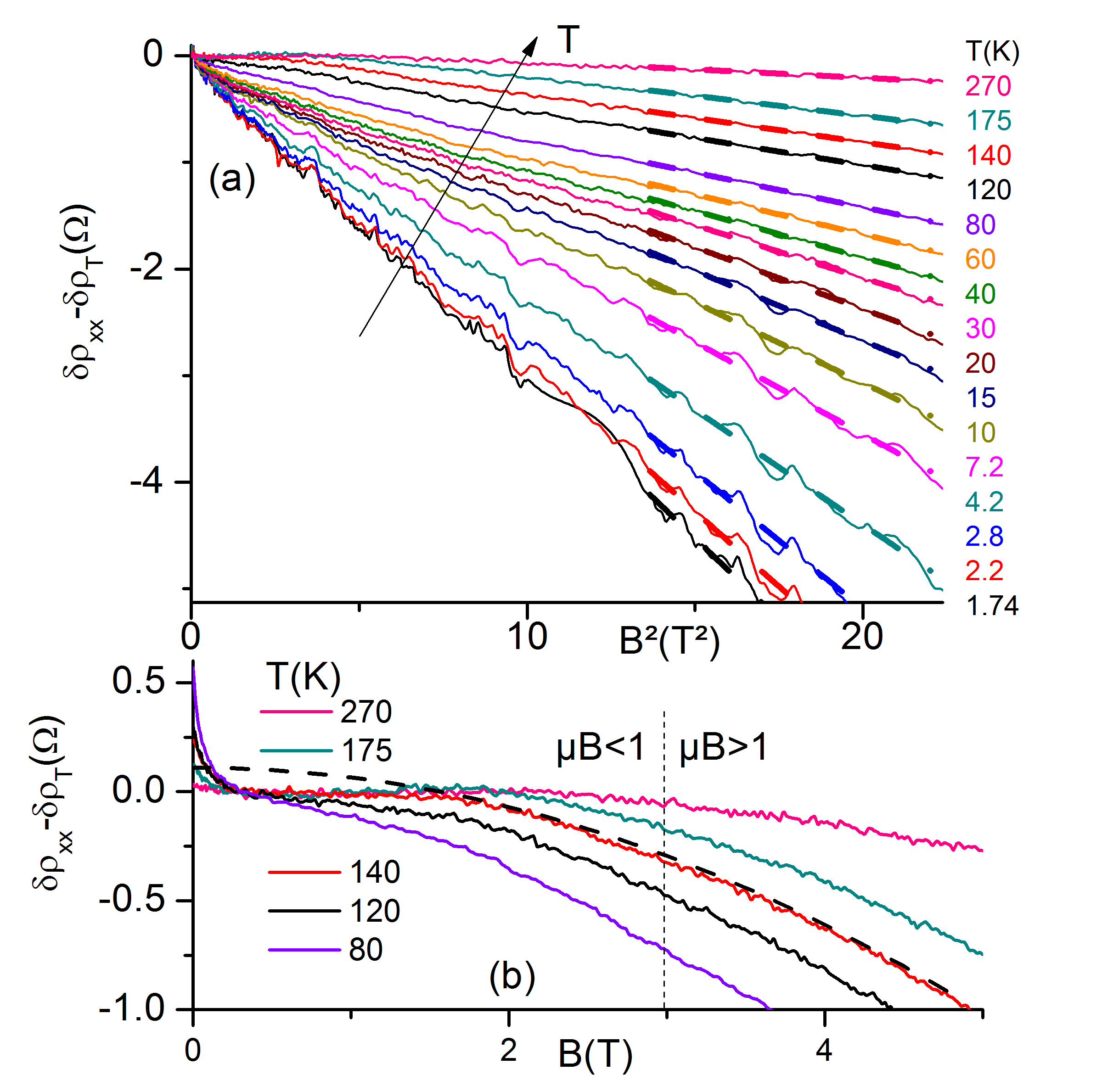}
\caption{(a) Solid lines: longitudinal resistivity $\delta\rho_{xx}$ {\it vs} $B^2$ minus the correction $\delta\rho_T$= $\rho_T(B)-\rho_T(0)$ due to thermal averaging, at different temperatures, from $T$=1.7 to 270 K. Dashed lines: best parabolic fits obtained in the strong $B$ regime. (b) Same data {\it vs} $B$ at  high $T$. The dashed line is a generic parabola used as guide for the eye to evidence the flattening at low $B$.}
\label{fig:rmt}
\end{figure}

\section{Discussion}
\subsection {Magnetoresistance maxima}
We first compare our results with previous investigations of MR maxima in other semiconductor structures. Kuntsevich {\it et al.} reported the existence of a similar maximum of the magnetoresistivity in the ballistic regime in various Si and GaAs two dimensional electron gases.\cite{Kuntsevich2009}  They showed that the maximum presents a universal behavior vs temperature, which we also retrieve for our samples: (i) it is a small effect, less than $1$ $\%$ ($0.5$ $\%$ in our case, which corresponds to a conductivity correction of the order of $G_0=e^2/\pi h$), (ii) it appears for not too-low temperatures $k_BT\tau/\hbar\geq1.3$ ($1.16$, in this work) and, (iii) the MR maximum grows and moves to higher field as $T$ increases, in accordance with Figs. \ref{fig:data}(c). Furthermore, the authors pointed out discrepancies between their experimental results and the available theory of Sedrakyan and Raikh, which predicts a non-monotonic behavior of the magnetoresistance.\cite{Sedrakyan2008} This theory predicts a $T$-independent maximum in the magnetoresistivity at $\mu B_{max}=1/\sqrt{3}$, and a decrease of the amplitude of the maxima when $T$ increases. These predictions do not correspond to our experimental situation, see Fig.~\ref{fig:data}(c).  
Quasiclassical memory effects are also known to lead to strong PMR in the presence of smooth long-range disorder or mixed disorder.\cite{Mirlin1999,Polyakov2001} However, this effect 
is temperature independent.
Finally, a MR maximum was recently detected in epitaxial graphene, see Fig.~4 of Ref.~\cite{Iagallo2013} but its interpretation in terms of EEI corrections led to anomalous values of the interaction parameter.%
%
%
%
\begin{figure}
\includegraphics[width=1.0\columnwidth]{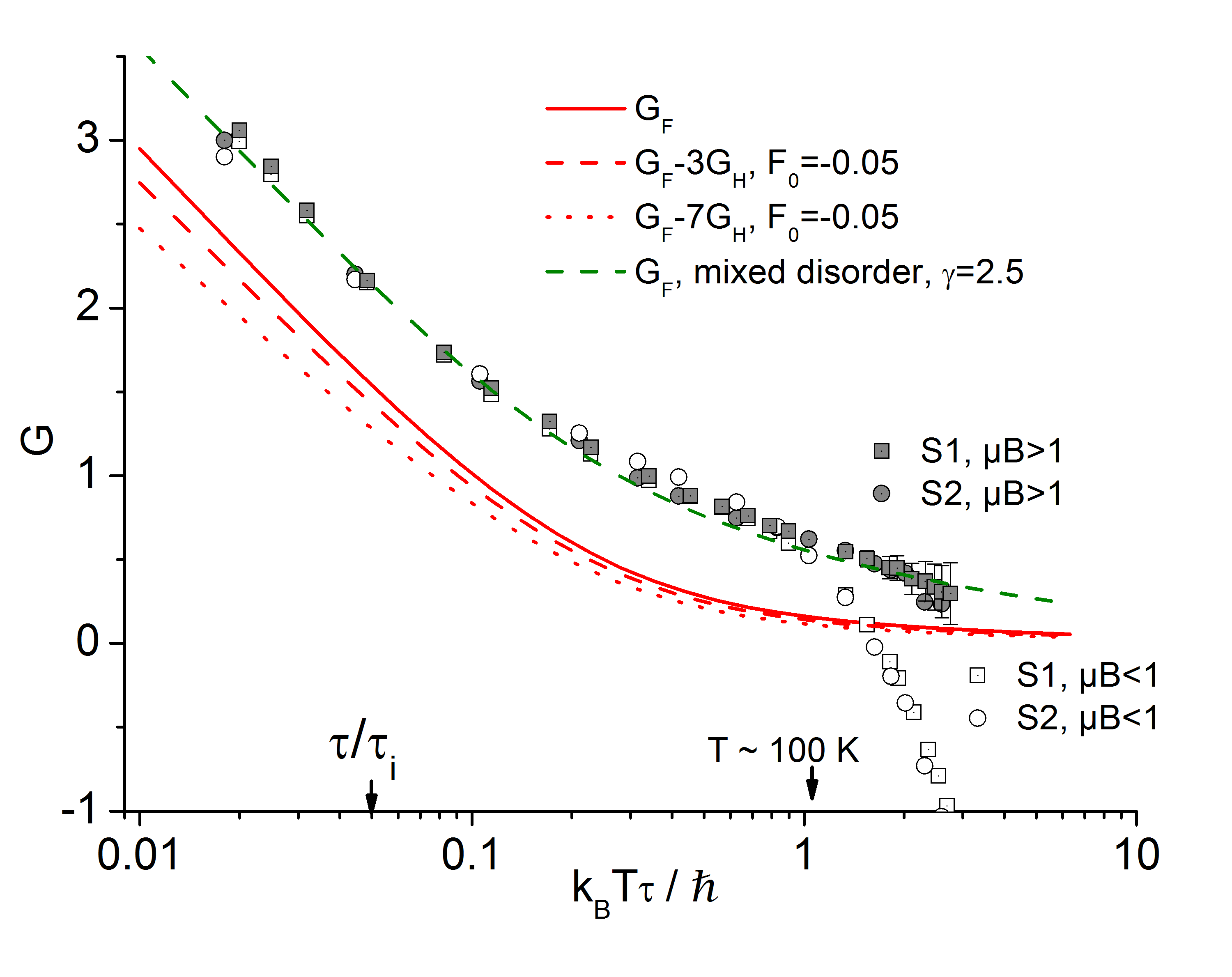}
\caption{Curvature of the magnetoresistivity {\it vs} $k_B T\tau/\hbar $. Solid squares and circles correspond to the MR at strong $B$ for samples S1 and S2. Open squares and circles are the MR at low $B$ ($\mu B <$ 1) for the same samples, without the subtraction of $\rho_T$.
The solid, dashed and dotted red solid lines are the theoretical fits $G_F$, $G_F-3G_H$ and $G_F-7G_H$ as calculated from Ref.~\cite{Gornyi2004}. The dashed green line corresponds to a situation of mixed disorder.}
\label{fig:g} 
\end{figure}

\subsection{PMR and thermal averaging}
We attribute the resistance maximum to a competition between
a PMR at low field induced by energy averaging within the temperature window around $\epsilon_F$, 
and the NMR due to EEI which persists at high fields.
In the framework of the relaxation time approximation, 
the conductivities $\sigma_{xx}$ and $\sigma_{xy}$ are given by:\cite{dassarmareview, Alekseev2014}
\begin{equation}
\sigma_{xx}= \frac{p e} {\langle 1 \rangle}
\left\langle \frac{\mu}{1+\mu^2 B^2} \right\rangle, ~~
\sigma_{xy}= \frac{p e B} {\langle 1 \rangle}
\left\langle \frac{\mu^2}{1+\mu^2 B^2} \right\rangle,
\label{eq:sxxsxy}
\end{equation}
where the brackets correspond to
\begin{equation}
\langle A \rangle= \frac{1}{\pi \hbar^2 v_F^2}
\int 
\left(-\frac{\partial \eta_F(\epsilon)}{\partial \epsilon}\right)
 A(\epsilon) \epsilon^2 d\epsilon, 
\end{equation}
$\epsilon$ is the energy with respect to the Dirac point, 
$\eta_F$ is the Fermi distribution function.
The magnetoresistance is then calculated as
$
\rho_{T}(B)= \sigma_{xx}/(\sigma_{xx}^2+ \sigma_{xy}^2).
$
This expression will give rise to PMR if $\mu$
cannot be taken out of the brackets in the above equations, 
{\it{i.e.}} if $\mu$ depends on $\epsilon$.
Then, for our experimental situation $T \ll \epsilon_F$ and at low magnetic fields, 
$\delta \rho_{T}/ \rho_{T} \sim \mu^2 B^2 (k_B T/\epsilon_F)^2$,
where the factor $(k_B T/\epsilon_F)^2 \ll 1$ arises due to the  weak
$\epsilon$-dependence of the mobility near the Fermi energy.~\cite{Alekseev2014}

To show this on a more formal level, we model the PMR by introducing 
scattering by phonons and ionized impurities.
Graphene phonon scattering is given by~\cite{Hwang2008}:
\begin{equation}
\tau_{ph}^{-1} =
\frac{\epsilon}{4\hbar^3 v_F^2}\frac{D^2}{\rho_m v_{ph}^2} k_B T,
\label{eq:tauph}
\end{equation}
where $D= 18.5$ eV is the expected deformation-potential coupling constant,~\cite{Hwang2008, Kozikov2010} $\rho_m = 7.6 \times 10^{-8}$ g$\cdot$cm$^{-2}$ is
the graphene mass density and $v_{ph}= 2 \times 10^{4}$ m/s the phonon velocity.~\cite{Hwang2008}
Ionized impurity scattering can be calculated within the Thomas-Fermi approximation:~\cite{dassarmareview}
\begin{equation}
\tau_{ii}^{-1} =
\pi \alpha^2 v_F^2 \hbar n_i \epsilon
\int_{-\pi}^{\pi}
\frac{ (1-\cos{\theta}^2)}
{\left(2 \epsilon \sin(\theta/2) + 4 \alpha \epsilon_F\right)^2}d\theta.
\label{eq:tauii}
\end{equation}
Here, $\alpha=e^2/\hbar v_F \bar{\kappa}$ is the interaction parameter, $\bar{\kappa}$ is the averaged dielectric constant of the environment,
$n_i$ the concentration of ionized impurities. 
The mobility $\mu= e\tau/m$, where $m$ is the cyclotron mass, is calculated by using the Matthiessen rule:
$
\tau^{-1} = \tau_{ii}^{-1} + \tau_{ph}^{-1}.
$
Within this model,  the experimental temperature dependence of the resistivity is well reproduced between 1.7 K and 150 K, as indicated by the fit reported in the inset of Fig.~\ref{fig:data}(a). The deviation from the fit remains small at higher temperatures and does not exceed 6\%.
The overall weak and linear increase of the resistivity with $T$
confirms that the graphene is well decoupled from the phonon modes of the interface, as expected when the SiC interface has been hydrogenated.~\cite{Jabakhanji2014}
Fig.~\ref{fig:data}(b) shows that the correction $\delta\rho_T= \rho_T-\rho_T(B=0)$ can also fit satisfactorily the amplitude of the PMR, as well as its slope at low $B$.
A direct estimate of $\alpha$ from the theoretical dielectric constants of SiC ($\kappa$ =9.66) and PMMA  ($\kappa \sim 5$) leads to $\alpha \sim 0.3$. This gives good fits with $n_i$ as the fitting parameter. Nevertheless, this overestimates the PMR by 25\% at room temperature.
To get an even better agreement we used both $n_i$ and $\alpha$ as fitting parameters.
The best fits shown in Fig.~\ref{fig:data}(a) and (b) have been obtained for $n_i \sim p$ and $\alpha = 0.16$.
The largest part of the PMR comes from the screening term $4 \alpha \epsilon_F$ in $\tau_{ii}$. If this term is neglected, there is no PMR induced by ionized impurities and the model fails as the PMR induced by phonon scattering alone is too small.
Other mechanisms have been considered: polar optical phonons from the PMMA~\cite{Fratini2008} give a very small PMR. Very little is known on the phonons of the SiC/graphene hydrogenated interface. Phonons of the non-hydrogenated SiC/graphene interface give, by a deformation potential,  a dependence ($\tau \sim \epsilon/T$)~\cite{Ray2012}, which suggests that in the case of hydrogenated interface, this mechanism does not contribute to the magnetoresistance.
Short range and resonant scatterers have not been included for simplicity in the model. They give a relative PMR five times larger than the one observed. 
Combining all this, ionized impurities are one of the most probable sources of the observed PMR.

\subsection{NMR and EEI correction}

We can now extract the EEI correction at all temperatures, by subtracting the term $\rho_T$ from the magnetoresistivity $\rho_{xx}$, see Fig.~\ref{fig:rmt}.
At $k_BT\tau/\hbar > 1$ ($T>100$ K) and $\mu B<1$ ($B<3$ T), the corrected curves become flat, as shown in panel (b). 
This is one of the key features predicted in Ref.~\cite{Gornyi2004}:  when the ballistic regime is approached and the disorder is smooth, the parabolic EEI corrections are strongly suppressed at $\mu B < 1$. 
For $\mu B > 1$, EEI correction is preserved and all curves can be fitted by parabola. The fits are shown as thick dashed lines and their slopes give the dimensionless curvature $G$:
\begin{equation}
\left( \delta\rho_{xx}- \delta \rho_T \right) /{\rho_{0}^2}=
-{(\mu B)^2} G_0 G .
\label{eq:curv}
\end{equation}
The curvature $G$ extracted from Fig.~\ref{fig:rmt} is plotted in Fig.~\ref{fig:g} for samples S1. The curvature of S2 is also indicated for comparison.
The thermal correction $\delta\rho_T$ plays a role only above $k_BT\tau/\hbar >2$. It then introduces uncertainties 
which are reported with error bars for sample S1. 
The raw MR coefficient at $\mu B<1$, without subtracting $\rho_T$, is also plotted.
The two coefficients at low and strong fields 
coincide at $k_B T\tau/\hbar < 1$ and diverge only at higher $T$.

The theoretical curvature is given by $G_F-cG_H$,
where $G_F$ is the exchange contribution, $G_H$ the Hartree contribution and $c$ the number of multiplet channels participating to the EEI. $G_F$ and $G_H$ are calculated following Ref.~\cite{Gornyi2004}.
The positive $cG_H$ term  
can only reduce both the curvature and the slope of the bare $G_F$ term.
The exchange term $G_F$ only, as plotted in Fig.~\ref{fig:g}, is already below the data points,
with a slope in the diffusive regime similar to the experimental one.
Therefore the Hartree term $c G_H$ is too small to be detected in our experiments.
This conclusion is reinforced by the constant slope $dG/dT$ observed around 
$k_B T\tau/\hbar  \sim \tau/\tau_i$.
In graphene $c$ is expected to change from 3 to 7 when $k_B T\tau/\hbar$ exceeds $\tau/\tau_i$ and this should modify the curvature.~\cite{Kozikov2010, Jobst2012}
The corrections expected for $G_F-3G_H$ and $G_F-7G_H$ and $F_0=-0.05$ are plotted in Fig.~\ref{fig:g}.
$G_H$ depends on the Fermi liquid constant $F_0$ which is estimated by the formula:
$
F_0= -{\alpha} \int_0^\pi \frac{\cos(\theta/2)^2 }
{ (\sin(\theta/2)+2 \alpha)} d\theta/2\pi.
$
While this formula was first proposed in Ref.~\cite{Kozikov2010} to explain low values of the experimental $F_0$,  it could still overestimate $F_0$. 
Indeed, it gives $F_0 \sim -0.05$ for $\alpha=0.16$, a value which 
gives too small slopes $dG/dT$ with respect to the experimental data, see Fig.~\ref{fig:g}.  
The change of slope at $k_B T \tau/\hbar \sim 0.3$ is therefore attributed to the transition from the diffusive to the ballistic regime.

We comment now on the vertical shift $\delta G \sim 0.5 G_0$ observed in Fig.~\ref{fig:g} between experiment and $G_F$. 
In the case of mixed long range and short range disorders, the interaction induced curvature $G_F$ is increased
by a factor $\alpha G_1 /2$, where $G_1$ is defined in Ref.~\cite{Gornyi2004} and $\alpha$ is a prefactor
which we take equal to $1-3/4\gamma-1/4\sqrt{\gamma}$,~\cite{note2}
$\gamma =(\tau_{sm}/\tau)$, $\tau_{sm}$ is the transport relaxation time of the smooth disorder only. 
From the best fit (green curve), we get $\gamma \sim 2.5$. We assume that $\tau^{-1} = 1/\tau_{sm}+1/2\tau_{wn}$, where $\tau_{wn}$ is the mean free time due to short range white noise potential and the factor 2 comes from the suppression of the backscattering due to the conservation of the pseudospin. This gives $2 \tau_{wn} \sim$ 100 fs with a corresponding length $l_{wn}$ = 50 nm.  

Another mechanism of classical origin can also increase the MR curvature~\cite{Mirlin2001} and gives the observed amplitude of the observed MR for a mixed disorder situation.~\cite{Li2003} 
However, this mechanism is temperature independent while experimentally $\delta G$ decreases at high temperature, see Fig.~\ref{fig:g}.

\subsection{Raman}
\begin{figure}
\includegraphics[width=0.9\columnwidth]{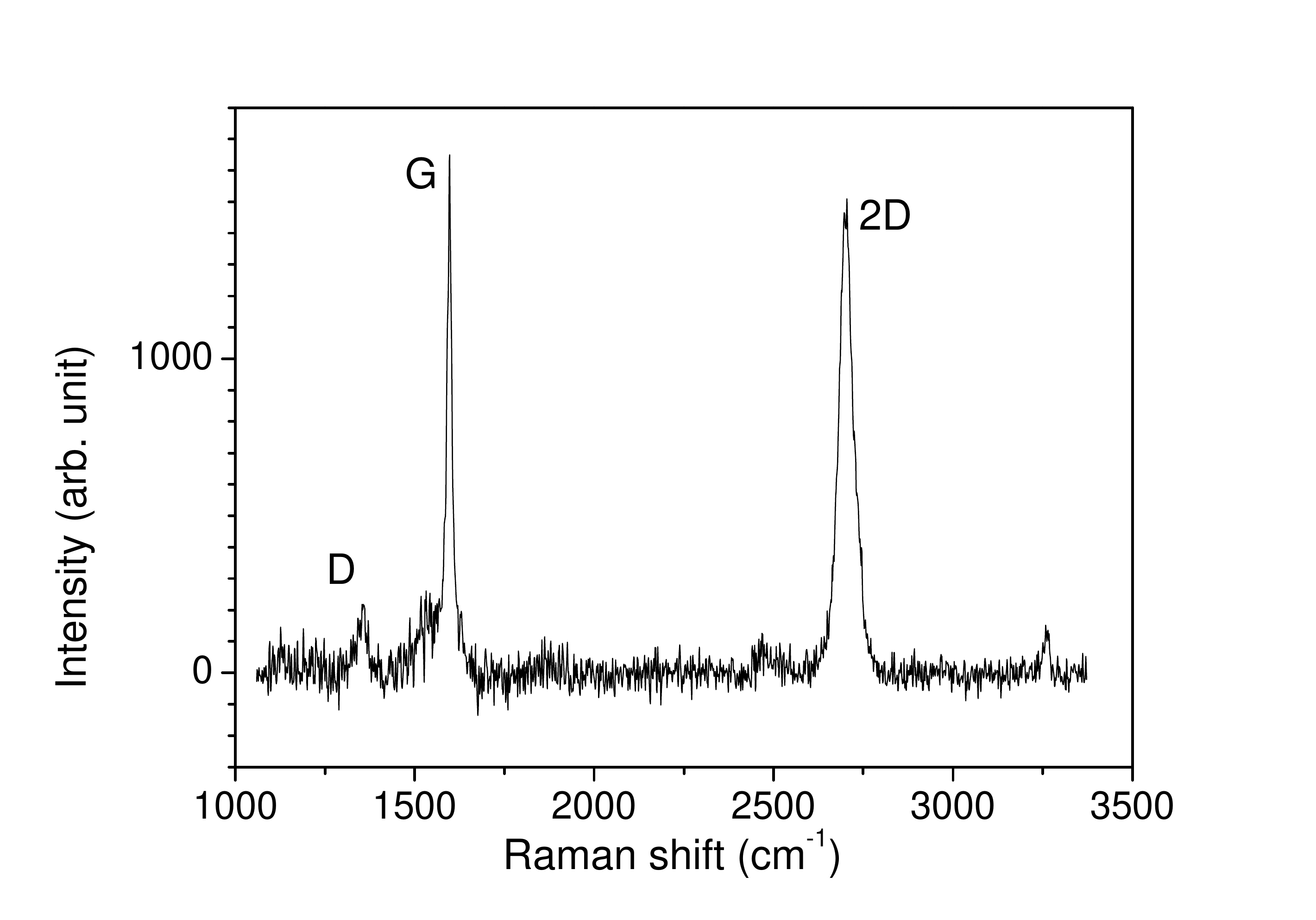}
\caption{Raman spectrum for sample S1. }
\label{fig:raman} 
\end{figure}

This distance $l_{wn}$ is comparable to the distance $l_d$ between structural defects extracted from Raman spectroscopy.
A typical Raman spectrum for sample S1 is presented in Fig.~\ref{fig:raman}.
To estimate the mean distance $l_d$ between the defects, we can calculate the intensity ratio of the $D$ and $G$ peaks which are observed in Raman spectroscopy. 
We then use the formula~\cite{cancado}: 
\begin{equation}
\left\{ l_d^2 \right\}_{\mathrm{nm}^2} = \frac{4300}{\{E_l\}_{\mathrm{eV}}^4}
\left( \frac{I_G}{I_D}
\right)
\label{eq:pim}
\end{equation}
where $\{E_l\}_{\mathrm{eV}}$ is the energy of the laser beam in eV, $I_D$ and $I_G$ are the integrated intensities of the $D$ and $G$ bands respectively. The ratio $I_G/I_D$ is $\sim 8 \pm 0.5$, which corresponds to an average distance $l_d \sim 35 $ nm.  \textbf{}

\subsection{More extensive comparison with the litterature}
\begin{figure}
\includegraphics[width=0.9\columnwidth]{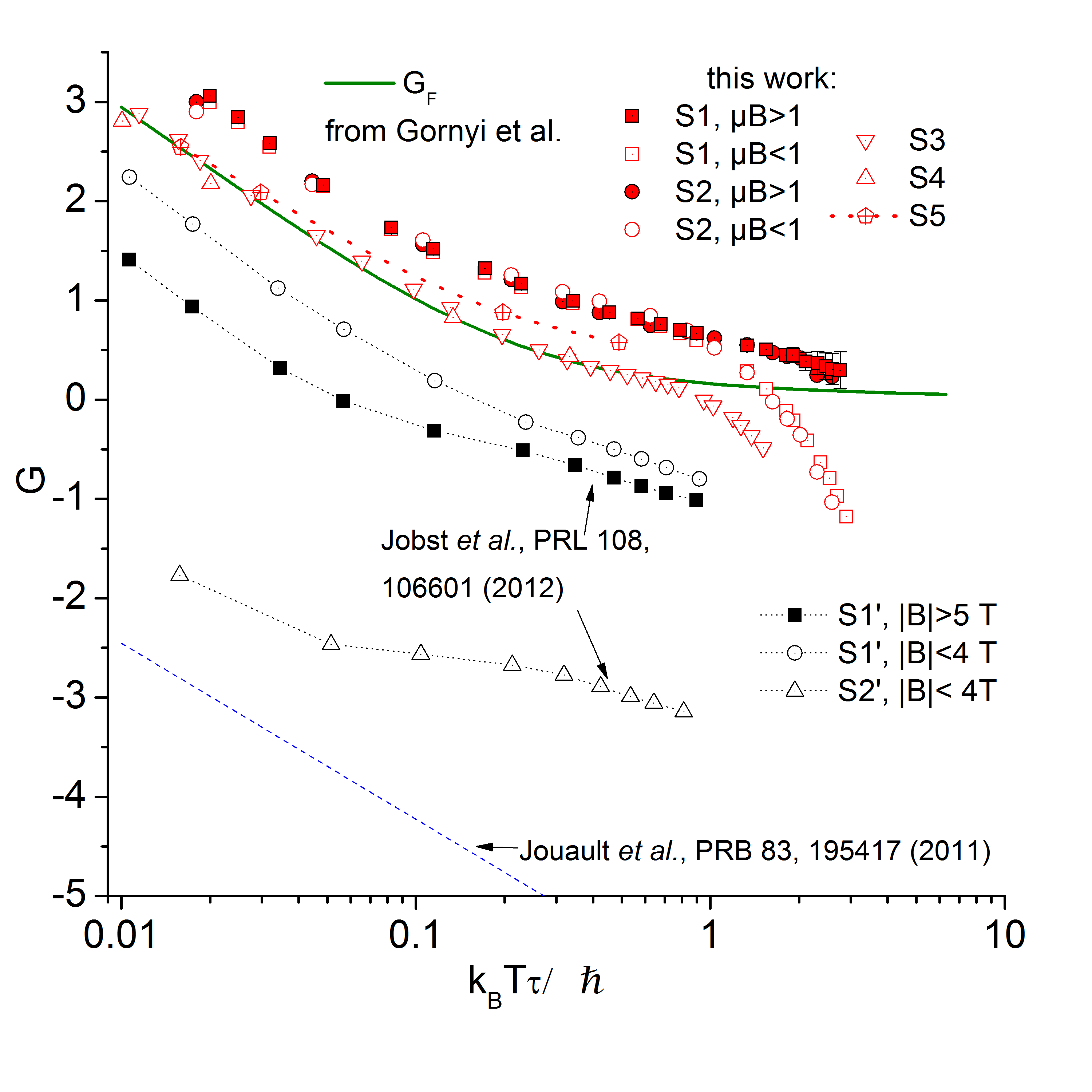}
\caption{Magnetoresistivity curvature {\it vs} $k_B T \tau/\hbar $. 
As in Fig.~\ref{fig:g}, solid squares and circles correspond to the MR curvature at $\mu B > 1$ for samples S1 and S2. The term $\delta \rho_T$ introduced in the main text is taken into account. 
Open squares and circles are the MR at $\mu B < 1$ for the same samples, calculated with $\rho_T$ neglected. 
Open triangles and diamonds plot the raw MR curvature at $\mu B < 1$ of three other samples S3, S4 and S5, with $\rho_T $ neglected. The green solid line is the theoretical fit $G_F$ from Ref.~\cite{Gornyi2004}. Data curves from Refs.~\cite{Jobst2012} and~\cite{Jouault2011} are also indicated.}
\label{fig:g2} 
\end{figure}
The magnetoresistances of the five samples S1-S5 have been studied and 
their curvatures are reported in Fig.~\ref{fig:g2}. 
All samples have $\tau / \tau_q \sim 4$, with the exception of sample S3, for which $\tau / \tau_q \sim 2$. 
For samples S3, thermal averaging completely dominates the EEI correction at room temperature, on the whole magnetic field range. Samples S4 and S5 have been studied up to $T \sim 100$ K only.  
A difference in the curvature below and above $\mu B=1$ could be detected only for samples S1 and S2. For comparison, additional data taken from Ref.~\cite{Jobst2012} and~\cite{Jouault2011} are also reported in Fig.~\ref{fig:g2}. The curve from Ref.~\cite{Jouault2011} corresponds to an average of the two main samples studied in this reference.

All the curves in Fig.~\ref{fig:g2} are approximately identical. 
First, with the exception of the curve from~\cite{Jouault2011}, all curves in Fig.~\ref{fig:g2} have a change of their slopes around $k_B T\tau/\hbar \sim 0.2$. For our samples, we attribute this change to the transition from the diffusive to the ballistic regime, as predicted by the theoretical $G_F$ curve.

Second, in the diffusive regime, the curves have the same slope for $G(\ln(T))$. However, they correspond to samples with different dielectric environment. In particular, samples from Ref.~\cite{Jouault2011} are not covered by PMMA, while the samples of this work are. This should lead to variations of the slope via the Hartree term $G_H$. Therefore, the influence of $G_H$ is probably too small to be detected. This in turn reinforces the attribution of the slope change to the diffusive-ballistic transition, and not to a modification of the number $c$ of multiplet channels participating to EEI.

Third, the curves differ mainly from the theoretical expectation $G_F$ by a vertical shift. Curves shifted downwards are probably prone to parasitic positive magnetoresistance induced for instance by current deflection or improper geometry. Curves shifted upwards are more intriguing. The enhanced negative curvature may result from the additional presence of short-range scatterers, as discussed previously.The very good agreement between $S3$ and the theory is possibly fortuitous, as both parasitic PMR and NMR can be present and compensate for each other.

\section{conclusion}
In conclusion, the magnetoresistance of monolayer graphene has been studied from 1.7~K to room temperature. The MR is well described by a recent theory of EEI valid for both diffusive and ballistic regimes. The overall enhanced negative curvature of the magnetoresistance points toward a situation of mixed disorder. This observation is sustained by additional Raman analysis. For graphene on SiC, the dominant scattering probably depends on the quality of the substrate. Finally, the dominance of short range scattering in our samples is in accordance with recent publications, where it is found that the mobility increases at low carrier density for epitaxial graphene.~\cite{Farmer2011, Satrapinsky2013}

\section{Acknowledgement}
We thank I. Gornyi for enlightening discussions.
This work was partly supported by the ANR project MetroGraph ANR-2011-NANO-004-06.


\begin{thebibliography}{43}
\bibitem[{not()}]{note2}
\bibinfo{note}{The value of $\alpha$ has been determined to adapt the formula
  proposed in Ref. 22 to low values of $\gamma$.}
\end{thebibliography}

\section{Appendix: fits of the SdH oscillations}
We fitted the SdH oscillations of the samples using the Lifshitz-Kosevich~\cite{lk, PhysRevB.71.125124} formula in which only the first harmonic is retained: 
\begin{equation}
\delta R_{xx}  \propto D_T D_D \cos( j \pi  \epsilon_F/\hbar \omega_c - \varphi ).
\label{lk}
\end{equation}
Here  $D_D$ is the Dingle factor:
$D_D=  \exp(-\pi/\omega_c  \tau_q)$,
$D_T$ is temperature amplitude factor: 
$D_T = \gamma/ \ sinh(\gamma)$ with $\gamma$=$2\pi^2k_BT/\hbar\omega_c$;
$\tau_q$ is the quantum time, $E_F$ is the Fermi energy, 
$\omega_c= e B/ m_c$ is the cyclotron frequency, $m_c$ is the cyclotron mass,
$\varphi$ is a phase factor and $j$ an integer.
The phase $\varphi$ determines the nature of the carriers. 
For a graphene layer:  $j=1$ and $\varphi=0$. 
For two dimensional massive carriers: $j=2$ and  $\varphi=\pi$.
For three dimensional massive carriers, $j=2$ and  $\varphi \approx 0.75 \pi $.
\begin{figure}[b]
\includegraphics[width=0.9\columnwidth]{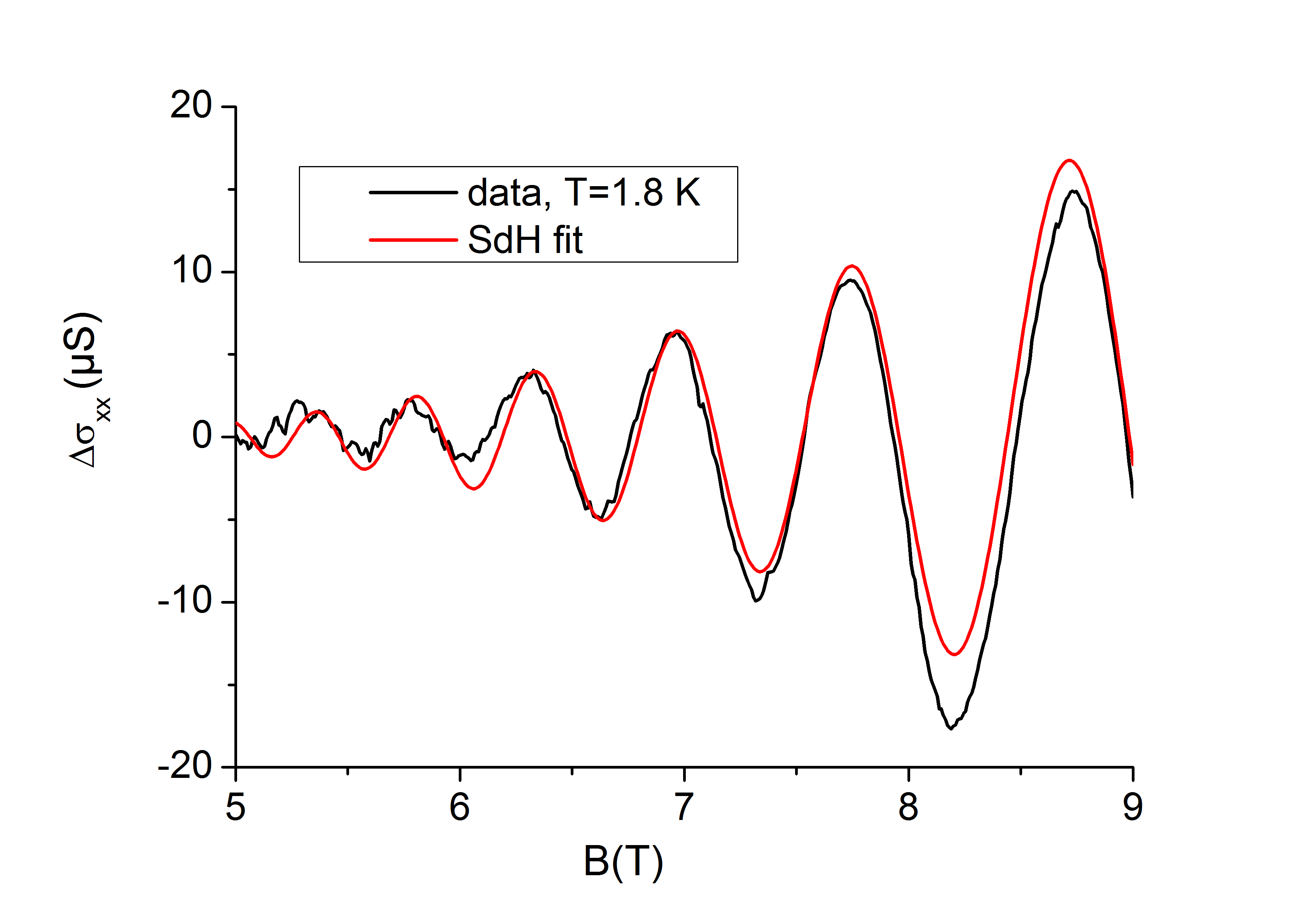}
\caption{Fit of the SdH oscillations at $T=1.8$ K for sample S1, from which the quantum time $\tau_q$ is extracted.}
\label{fig:sdhfit} 
\end{figure}
The best fit for sample $S1$ at $T$= 1.7 K is reported in Fig.~\ref{fig:sdhfit}.
It gives $\varphi \sim 0$ and $\tau_q \sim$ 17 fs.
\end{document}